\documentclass[letterpaper,twocolumn,10pt]{article}
\usepackage{usenix}

\usepackage{tikz}
\usepackage{amsmath}
\usepackage{hyperref}
\usepackage{graphicx}
\usepackage{booktabs}
\usepackage{tabularx}
\usepackage{array}
\usepackage{float}

\usepackage[table]{xcolor}
\usepackage{array}
\usepackage{tabularx}
\usepackage{caption}
\usepackage[most]{tcolorbox}
\usepackage{fontawesome5}

\usepackage{enumitem,amssymb}
\newlist{todolist}{itemize}{2}
\setlist[todolist]{label=$\square$}

\newtcolorbox{takeaway}{
    enhanced,
    breakable,
    width=0.98\columnwidth,
    colback=gray!10,
    colframe=black,
    boxrule=1pt,
    arc=2mm,
    left=2mm,
    right=2mm,
    top=1.5mm,
    bottom=1.5mm,
    before skip=6pt,
    after skip=6pt,
    fontupper=\small,
    before upper={
        \setlength{\parindent}{0pt}
        \faWpexplorer \hspace{0.3em}\textbf{Takeaway:}\space
    }
}

\newcounter{survey}
\newcommand{\surveyitem}{\stepcounter{survey}\item[\textbf{Q\thesurvey.}]}

\begin{document}

\date{}

\title{\Large \bf Understanding the Usability of Cryptographic Verification Tools}

 \author{
 {\rm Tarikul Islam}\\
 Samsung R\&D Institute Bangladesh
 \and
 {\rm Yasin Islam}\\
 Bangladesh University of Engineering and Technology
 \and 
{\rm Khandakar Ashrafi Akbar}\\
 Towson University
 \and
{\rm Imtiaz Karim}\\
 The University of Texas at Dallas
}

\maketitle

\begin{abstract}
Cryptographic protocol verification tools are widely used to analyze the security of complex protocols, yet how users interact with these tools remains comparatively understudied. We present an exploratory human-centered study of experienced users of Tamarin, ProVerif, and related protocol verifiers. Our survey included researchers, graduate students, and practitioners with hands-on experience using Tamarin, ProVerif, or related tools. The findings reveal usability barriers across the verification workflow, including difficulties debugging non-termination and performance issues and the lack of systematic methods for validating formal models against real protocols. When proofs fail without concrete attacks, users commonly simplify models, add helper lemmas, and revisit modeling abstractions. Participants also called for actionable diagnostics, clearer explanations of results, visualization, and automation for recurring proof tasks. Our findings suggest that persistent usability challenges arise from the gap between protocol-level reasoning and the verifier’s formal model, proof procedures, and diagnostic output. We derive concrete design priorities for improving the accessibility, interpretability, and usability of cryptographic protocol verification tools.
\end{abstract}

\section{Introduction}

Formal verification has become a cornerstone of modern protocol
security engineering. Tools such as ProVerif~\cite{blanchet_proverif, blanchet2024proverif} and the Tamarin
Prover~\cite{meier2013tamarin, basinTamarin} allow researchers and practitioners to symbolically model
cryptographic protocols, specify adversarial capabilities, and
either prove that a protocol satisfies a given security property for
an unbounded number of sessions or produce a concrete attack trace
that violates it. These tools have moved well beyond academic
exercises: Tamarin was used in the design and analysis of Apple's
iMessage PQ3 protocol~\cite{apple_pq3_2024} and helped uncover a security flaw in an early
draft of TLS~1.3~\cite{cremers2016automated}, while ProVerif has been applied to widely deployed
protocols including TLS 1.3~\cite{BhargavanBlanchetKobeissiSP2017} and Signal~\cite{blanchet2012automatic, Kobeissi2017}. As protocols
grow more complex, incorporating features such as post-compromise
security, ratcheting, and stateful hardware components, the
practical demand for rigorous, tool-assisted verification continues
to grow.

Yet the effectiveness of a verification tool depends not only on its
theoretical expressiveness but on whether its intended users can
actually apply it correctly. A protocol verifier that can, in
principle, establish strong security guarantees is of limited value
if practitioners cannot construct a faithful model of the protocol,
cannot understand why a proof fails to terminate, or cannot
distinguish a genuine attack from an artifact of an overly coarse
abstraction. Usable security research has long shown that this kind
of gap between theoretical capability and practical usability is a
recurring source of real-world security failure: studies of
cryptographic APIs, for instance, have found that developers can
produce code they \emph{believe} is secure while violating basic
security requirements, not because the underlying primitives are
flawed, but because the tools and interfaces around them are hard to
use correctly~\cite{acar17, nadi16}. Adjacent classes of security tooling, including
constant-time analysis tools and fuzzers, exhibit similar patterns:
usability shortcomings around installation, documentation, and
output interpretation measurably suppress adoption even when the
underlying technique is mature and effective~\cite{fourne2024these, zhao2025qualitative}.

Despite this, the usability of cryptographic protocol verification
tools themselves has received comparatively little systematic
attention. Prior work on Tamarin and ProVerif overwhelmingly focuses
on their formal expressiveness and analytical power, for example
comparing which tool can detect which class of attack on a shared
benchmark of protocols~\cite{hassan2025evaluating}. This line of work is valuable for
understanding \emph{what} these tools can prove, but it says little
about \emph{how} real users, ranging from graduate students
encountering formal methods for the first time to expert researchers
modeling novel protocols, actually experience the process of
learning the tool, constructing a model they trust, and interpreting
the tool's output. The formal methods community has itself begun to
recognize this gap: recent position papers call explicitly for more
human-centered study of formal verification tools, arguing that
questions of learnability, mental models, and practical impact
remain underexplored relative to the pace of technical advances in
the tools themselves~\cite{krishnamurthi2019human, carreira2021exploring}. Industrial user studies of verification
engineers echo this concern, finding that the interpretability of
verification results, rather than the underlying formal guarantees,
is often what determines whether engineers trust and adopt a tool~\cite{kaleeswaran2023user}.

We address this gap and present a survey-based empirical study of 16 experienced users of cryptographic protocol verification tools. The users had direct, hands-on experience using Tamarin,
ProVerif, or related tools in academic or industry
settings, recruited through academic publications, open-source
contributions, community outreach, and snowball sampling. The
survey combined closed- and open-ended questions across five
sections, covering participant background, tool learnability, the
end-to-end modeling and verification workflow, and tool-specific
usability issues unique to Tamarin and ProVerif respectively.
Guided by this study, we investigate three research questions:

\begin{itemize}
    \item \textbf{RQ1:} What usability barriers do users encounter when applying cryptographic protocol verification tools such as Tamarin and ProVerif?
    \item \textbf{RQ2:} How do users perceive the transfer of prior programming and general purpose model-checking experience to learning and using cryptographic protocol verification tools?
    \item \textbf{RQ3:} What improvements do users believe would make these tools more accessible and effective?
\end{itemize}

Our findings show that usability barriers are distributed across the
entire verification workflow rather than concentrated at any single
stage. While learning a tool's syntax and underlying formal theory
poses an initial hurdle, participants consistently identified the
most persistent difficulties elsewhere: validating that a formal
model faithfully captures the real protocol, debugging proofs that
fail to terminate or fail without a concrete counterexample, and
interpreting sparse or non-actionable error messages and progress
output. We find that programming experience transfers partially,
supplying useful problem-solving strategies such as algorithmic
thinking and case analysis, but does not substitute for the
domain-specific reasoning that symbolic protocol verification
demands, and that prior exposure to general-purpose model checkers,
while helpful, is not a prerequisite that most users bring with
them. Tool-specific findings further show that Tamarin users
frequently rely on manually constructed helper lemmas and
interactive proof inspection to manage scalability, while ProVerif
users report persistent friction modeling stateful protocols,
complex algebraic theories, and advanced properties such as
post-compromise security. Across both tools, participants converge
on a consistent set of desired improvements: mechanisms for
validating models against the protocols they represent, more
actionable and localized error messages, interactive debugging and
visualization support, and automation for common but tedious tasks
such as helper-lemma generation.

This paper makes the following contributions:

\begin{itemize}
    \item We present an exploratory human-centered study focused specifically on the end-to-end experiences of 16 experienced users of Tamarin, ProVerif, and related cryptographic protocol verifiers.
    \item We identify and characterize usability barriers spanning
    the full verification workflow, from learnability through
    modeling, proof debugging, and result interpretation, and show
    that these barriers are shared across tools despite their
    differing theoretical foundations.
    \item We analyze the extent to which programming and
    general-purpose model-checking experience transfers to
    cryptographic protocol verification, providing evidence for
    where existing user knowledge helps and where it falls short.
    \item We distill participants' feedback into concrete,
    prioritized recommendations for improving the usability,
    diagnostics, and automation support of cryptographic
    verification tools, informing both future tool design and the
    development of educational and documentation resources.
\end{itemize}


\section{Background}
\looseness-1
Cryptographic protocol verification tools analyze formal models of protocols under explicitly defined adversarial capabilities.
Many such tools adopt the symbolic, or Dolev--Yao, model~\cite{dolev1983security}, in which cryptographic operations such as encryption and signatures are represented as symbolic functions rather than concrete bit-level computations. Cryptographic primitives are assumed to behave ideally: for example, an adversary can decrypt an encrypted message only when it possesses the required key. Verification therefore reasons about which symbolic terms an adversary can derive from its initial knowledge and from messages observed or generated during protocol execution. This abstraction enables automated reasoning about protocols with potentially unbounded executions while separating protocol-level reasoning from computational properties of the underlying cryptographic primitives. ProVerif and Tamarin are among the most widely adopted tools for performing this type of symbolic security protocol analysis.

\looseness-1
\noindent \textbf{Tamarin}~\cite{meier2013tamarin, basinTamarin} models protocol behavior using multiset rewriting rules and expresses security properties using first-order logic. Its symbolic reasoning supports equational theories and protocols involving mutable state and temporal relationships between events. Tamarin provides both automated and interactive verification. Automated proof search can either establish a property or produce a counterexample trace, but termination is not guaranteed. In interactive mode, users can inspect intermediate proof states and dependency graphs and manually guide the search, including through auxiliary or helper lemmas. These characteristics make proof guidance, non-termination, trace interpretation, and stateful modeling particularly relevant to Tamarin's usability.

\looseness-1
\noindent \textbf{ProVerif}~\cite{blanchet2001cient, blanchet_proverif, blanchet2024proverif} is an automated symbolic protocol verifier based on the applied pi calculus~\cite{piCalculus, abadi2001mobile}. It translates protocol models into an abstraction that enables reasoning about an unbounded number of protocol sessions and an unbounded message space. The tool supports a range of cryptographic operations through equational and functional theories and can analyze properties including secrecy, authentication, and forms of equivalence. ProVerif's verification engine primarily emphasizes automated analysis. Although it also provides an interactive process simulator, this differs from Tamarin's workflow for interactively inspecting and guiding proof search. Its abstractions can make verification highly scalable, but the modeling of features such as mutable state, complex algebraic properties, equivalence properties, and state-evolving security guarantees may require additional modeling choices or workarounds.

These differences are important for our study. Both tools require users to translate an informal protocol into a symbolic representation and interpret the resulting verification outcome, but they expose different modeling formalisms and proof workflows. We therefore examine both usability challenges shared across cryptographic verifiers and tool-specific issues arising from Tamarin's interactive proof process and ProVerif's automated abstraction-based analysis.

\section{Related Work}

\noindent\textbf{Usability in Cryptographic Software and APIs.}
Prior work has shown that API design, abstraction, documentation, defaults, examples, and required security knowledge affect developers' ability to use cryptographic software securely~\cite{nadi16, acar17, patnaik2021don, mindermann2018usable}. Controlled studies further show that interface design can materially affect security and functional correctness, while programming experience alone does not prevent security-relevant API blind spots~\cite{acar17,oliveira2018}. Accordingly, researchers have advocated developer-centered APIs, safer defaults, higher-level abstractions, and contextual guidance that reduce the amount of cryptographic reasoning required from users~\cite{green2016developers,gorski2018developers,gorski2020listen,firouzi2024struggle}. Similar abstraction and usability problems have been observed in Ethereum cryptographic APIs and in the design and use of cryptographic libraries more broadly~\cite{zhang2024contracts,schmuser2025m,patnaik2019usability}. Systematization work, however, suggests that many proposed security-API usability recommendations still lack extensive empirical validation~\cite{patnaik2021don}.

\looseness-1
\noindent\textbf{Usability of Security Analysis Tools.}
Adjacent security-analysis tools exhibit similar problems. Studies of static analysis identify false positives, poor warning prioritization, inadequate explanations, and weak workflow integration as barriers to effective use~\cite{johnson2013,christakis2016developers,sadowski2018lessons,nguyen2022}. Developers also need warnings that explain the cause, security relevance, and possible remediation of a reported problem~\cite{smith2015,smith2019}. Usability evaluations consequently report difficulties with navigation, scalability, alert interpretation, prioritization, and remediation, while warning presentation and specificity can influence developers' ability to understand and act on results~\cite{smith2020,tahaei2021security,nachtigall2022large}. Comparable difficulties have been reported for fuzzing, including setup, configuration, workflow integration, progress interpretation, and failure diagnosis~\cite{ploger2021usability,ploger2023usability,zhao2025qualitative}, and for constant-time analysis tools, where installation, documentation, and output interpretation can hinder routine use~\cite{fourne2024these}. Broader studies of developer security practices likewise emphasize the importance of usable feedback and support for recovering from security-relevant mistakes~\cite{votipka2020understanding,gorski2020listen}. These findings suggest that cryptographic protocol verifiers may face related challenges, compounded by the additional demands of formal modeling, proof construction, and result interpretation.

\looseness-1
\noindent\textbf{Usability and Human Factors in Formal Verification.}
Usability concerns have also been identified specifically in formal security-protocol verification. Garcia and Modesti~\cite{garcia2017ide} observed that specialized specification languages and fragmented workflows make protocol-analysis tools difficult to use and proposed an IDE integrating higher-level protocol modeling with OFMC and ProVerif. Their subsequent evaluation found that integrated tooling and abstractions can assist users with limited formal-methods or cryptography backgrounds, while complexity, training, integration, and interpretation of verification results remain important obstacles~\cite{garcia2024practical}. Braghin et al.~\cite{braghin2025aprover} similarly identify the specialized languages of tools such as Tamarin and ProVerif as an expertise barrier and propose higher-level textual and graphical representations for these verification back ends.

Broader formal-methods research reinforces these concerns. Surveys of researchers and practitioners identify ease of use, training, skills, scalability, tool maturity, and workflow integration as persistent adoption challenges~\cite{gleirscher2020formal,garavel2020}. Krishnamurthi and Nelson~\cite{krishnamurthi2019human} argue for greater attention to the humans interacting with formal methods, while Carreira et al.~\cite{carreira2021exploring} call for research into users' mental models and understanding of formal-verification guarantees. Empirical studies also highlight difficulties interpreting formal notation and verification outcomes, debugging failed proofs, and maintaining proofs over time; explanatory support can improve users' understanding and acceptance of verification results~\cite{kaleeswaran2023user,mugnier2025impact}. Ter Beek and Ferrari~\cite{ter2022empirical} further note that empirical evaluations of formal-method tools remain comparatively rare and advocate greater use of usability and human-subject studies.

\looseness-1
Taken together, prior work establishes that effective security and formal-verification tools depend not only on analytical capability, but also on learnability, appropriate abstractions, actionable feedback, debugging support, and alignment with users' workflows. Existing work has recognized usability barriers in security-protocol verification and proposed interfaces intended to reduce them, but systematic empirical evidence about how experienced users interact with cryptographic protocol verifiers such as Tamarin and ProVerif remains limited. In particular, little is known about how users learn these tools, how prior programming and model-checking knowledge transfers to them, how they assess model fidelity, and how they diagnose and recover from unsuccessful or non-terminating verification attempts. Our study addresses these gaps by examining users' experiences across the end-to-end verification workflow.

\looseness-1
\section{Methodology}
In this section, we provide an overview of the methodology used in our questionnaire-based survey study. We also discuss recruitment and screening procedures, participant eligibility criteria, and data analysis. To examine the usability of cryptographic verification tools, we conducted an online survey with participants who had practical experience using these tools in academic or professional contexts. Our goal was to understand how users interact with cryptographic verification tools, the challenges they face during modeling and verification tasks, and the improvements they believe would make these tools more usable. For reference, the survey questionnaire is included in the Appendix~\ref{app:survey}.

\subsection{Study Design}
We designed the study as an online questionnaire-based survey focused on user experiences with cryptographic verification tools. Because these tools are highly specialized and require substantial technical knowledge, we targeted participants with direct hands-on experience rather than general familiarity with formal methods or security protocols.

The survey was intended to capture both closed-ended and open-ended responses. Closed-ended questions were used to collect information about the background of the participants, tool usage and common usability challenges. Open-ended questions allowed participants to describe their experiences in more detail, including difficulties they encountered, strategies they used to overcome those difficulties, and suggestions for improving cryptographic verification tools.

The study focused on practical usability rather than theoretical robustness of the tools. In particular, we were interested in users’ experiences with learning the tools, writing protocol models, understanding tool syntax and semantics, interpreting error messages or verification results, debugging failed proofs or models, and integrating these tools into broader research or development workflows.

\subsection{Recruitment and Eligibility Criteria} We used purposive recruitment to reach participants with hands-on experience using cryptographic protocol verification tools in academic or professional settings. Because these tools serve a relatively small and specialized community, we recruited through four complementary channels: academic publications involving protocol verification, public technical contributions such as protocol models and tool-related repositories, relevant online research communities, and professional referrals and snowball sampling. Authors and contributors were identified as potential participants based on evidence of relevant tool use rather than assumed expertise.
Recruitment was conducted from May to July 2026. We sent approximately 100 recruitment invitations by email to potential participants identified through the recruitment channels described
above. The invitations included a link to the Qualtrics survey, and follow-up reminder emails were sent during the recruitment period.
In total, 24 individuals started the survey. Of these, 16 completed the survey and were retained for analysis, while eight incomplete
responses were excluded. Thus, our final sample consisted of 16 participants.
Eligibility was based on self-reported prior experience with cryptographic protocol verification tools. On the consent page, prospective participants were informed that the study was intended for individuals with experience using such tools and were asked to confirm that they met this requirement before proceeding. Participants also reported the specific tools they had used and their level of experience in the demographic portion of the survey.

\subsection{Study Structure}
We conducted the study using an online survey implemented in
Qualtrics~\cite{qualtrics}. The survey was organized into five
sections: consent, participant background and tool experience,
learnability, modeling and verification, and tool-specific
usability questions. Conditional display logic was used so that
participants were shown Tamarin and ProVerif specific questions only when they reported prior experience with the corresponding tool. Appendix Figure~\ref{fig:study-procedure} shows the complete survey flow.

\noindent \textbf{Consent.}
The participants were first presented with a consent page describing the purpose of the study, the procedure, the potential risks and benefits, and the participation requirements. Only participants who selected ``I agree'' were allowed to continue.

\noindent \textbf{Participant background.}
Participants reported their current role, experience with
cryptographic verification tools, self-rated expertise, and the
specific tools they had used. Their tool selections determined
which tool-specific sections were subsequently displayed.

\looseness-1
\noindent \textbf{Learnability.}
This section examined how participants learned cryptographic
verification tools and the difficulties they encountered. Questions covered learning resources, prior knowledge, the steepest parts of the learning curve, transferable programming concepts, and previous experience with general-purpose model checkers.

\looseness-1
\noindent \textbf{Modeling and verification.}
Participants were asked about their confidence in model fidelity, factors limiting that confidence, strategies for handling unsuccessful proofs, communication during slow or non-terminating verification, and shortcomings in error messages.

\looseness-1
\noindent \textbf{Tool-specific usability.}
Participants with Tamarin experience received questions concerning
counterexample analysis, scalability, helper lemmas, dependency-graph
visualization, modeling difficulty, and desired improvements.
Participants with ProVerif experience received questions concerning
observational equivalence, algebraic properties, state-dependent
protocols, forward secrecy, post-compromise security, and desired
improvements.

The survey included both closed- and open-ended questions, allowing
participants to report specific difficulties, describe strategies
or workarounds, and suggest improvements. Table~\ref{tab:interview-protocol}
summarizes the survey structure, while Appendix~\ref{app:survey}
provides the complete questionnaire.
\subsection{Data Analysis}
\label{sec:data-analysis}
We analyzed closed-ended and open-ended survey responses using complementary quantitative and qualitative approaches.

\noindent \textbf{Quantitative analysis.}
For closed-ended and multiple-choice items, we report descriptive statistics (counts and percentages) across respondents. Because Qualtrics' conditional display logic showed tool-specific question blocks only to participants who reported experience with the corresponding tool, and because not all participants answered every item, we report the applicable $n$ for each question alongside its results.

\looseness-1
\noindent \textbf{Qualitative coding of open-ended justifications.}
Several survey items (e.g., Q20--Q24 in the ProVerif block) asked participants to rate the adequacy of ProVerif's support for a specific security property or protocol feature and to justify that rating in free text. To analyze these justifications systematically, we developed a codebook through an inductive, data-driven process. One researcher read all open-ended responses for a given question and derived a set of codes capturing the distinct concerns, caveats, or justifications participants raised, iteratively refining code definitions and splitting or merging codes as needed until they stably captured the range of responses. Each code was documented with a definition and one or more representative (de-identified) response excerpts, and annotated with the number of respondents assigned to it.

To enable comparison of perceived adequacy across the five properties (Q20--Q24), each per-question code was additionally mapped onto a shared five-level ordinal severity scale, ranging from \emph{Fully Adequate} (no significant caveats reported), through \emph{Adequate with Minor Limitations} and \emph{Adequate but Requires Significant Improvement} (verifiable in principle, but only via substantial manual effort, workarounds, or careful nonstandard modeling), to \emph{Insufficient} (a structural or formalism-level gap rather than a modeling inconvenience), and \emph{Not Sure} (no opinion expressed). This shared scale allowed us to aggregate and compare adequacy ratings across properties with different underlying codes, and underlies the response distributions reported in Appendix Tables~\ref{tab:q24}--\ref{tab:q20}.

Finally, we conducted a lightweight axial coding pass across the justifications for all five questions to identify cross-cutting themes that recurred across multiple properties rather than being specific to any single one, such as the manual effort required by workarounds, recurring termination concerns, and comparisons participants drew to Tamarin's support for algebraic reasoning. These cross-cutting themes inform the synthesis presented alongside individual question results in Section Results and the discussion in Section RQ3.

\noindent \textbf{Coding limitations.}
Open-ended responses were coded by a single researcher; we did not compute inter-rater reliability, and a second coder did not independently verify the coding. Given the relatively small number of free-text responses per question ($n=13$ per property) and the exploratory goals of the study, we treat the resulting codes and themes as a structured, transparent summary of participants' stated reasoning rather than as a statistically validated coding scheme. The full codebook, including per-question code definitions, severity-band mappings, and cross-cutting themes, is included in our open-science artifacts (see Open Science).

\begin{table}[t]
\centering
\caption{Summary of the survey instrument. See Appendix~\ref{app:survey}
for the complete questionnaire.}
\label{tab:interview-protocol}

\footnotesize
\renewcommand{\arraystretch}{1.38}
\setlength{\tabcolsep}{4pt}

\begin{tabularx}{\columnwidth}{
    @{}
    >{\raggedright\arraybackslash}p{0.36\columnwidth}
    >{\raggedright\arraybackslash}X
    @{}
}
\toprule
\textbf{Section (\# Questions)}
&
\textbf{Sample Questions}
\\
\midrule

Demographic (4)
&
Which of the following verification tools have you used?
\\[2pt]
\midrule

Learnability (4)
&
What resources did you rely on when learning the tool? (\emph{select all that apply})
\\[2pt]
\midrule

Modeling \& Verification (5)
&
How confident are you that your formal model accurately reflects
the real-world protocol you intend to verify?
\\[2pt]
\midrule

Tamarin (6)
&
What is the most important usability improvement you would like
to see in Tamarin Prover?
\\[2pt]
\midrule

ProVerif (6)
&
How adequate is ProVerif’s current support for reasoning about complex algebraic properties (\emph{e.g., Diffie–Hellman, XOR, group equations)? Please explain your reasoning for your choice.}
\\

\bottomrule
\end{tabularx}
\end{table}

\section{Results}
\subsection{Participants}
Our sample (see Appendix Table~\ref{tab:participant-demographics} for details) was predominantly experienced: 12 of 16 participants were academic researchers or faculty, two were graduate students, and two were industry professionals. Thirteen participants rated themselves as having Advanced or Deep Expertise, and nine reported more than six years of experience with security-protocol verification tools. Thirteen participants had used ProVerif, while twelve had used Tamarin, with several reporting experience across multiple verification tools.

These characteristics indicate that our findings primarily reflect the experiences of established users rather than novice users. This is important when interpreting the results: difficulties reported in later sections, particularly around model validation, proof debugging, and non-termination, persist even among participants with substantial prior experience.

\subsection{Learnability of Cryptographic Verifiers}
Before users can effectively apply cryptographic verifiers, they must learn both the tool-specific modeling language and the formal reasoning style that underlies the verification process. In this section, we examine how participants learned cryptographic verification tools, what resources they relied on, which parts of the learning process they found most difficult, and what prior programming knowledge helped them become more effective users. Similar to other specialized security tools, cryptographic verifiers require users to move beyond ordinary programming or testing workflows and adopt a more abstract way of representing protocols, adversarial behavior, and security properties. 

\begin{figure}[t]
    \centering
    \includegraphics[width=\columnwidth]{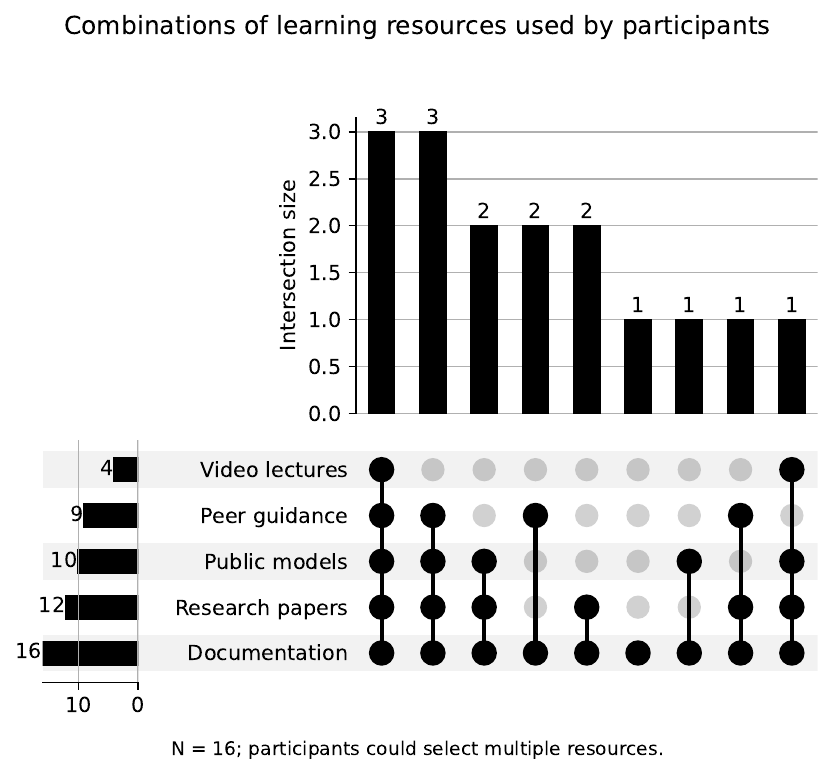}
    \caption{Combinations of learning resources used by participants. Filled circles identify the resources included in each combination, while the vertical bars show the number of participants reporting that combination.}
    \label{fig:learning-resources-upset}
\end{figure}

\noindent \textbf{Learning Resources. }Participants reported relying on a combination of official documentation, research papers, publicly available models, recorded lectures, and guidance from peers or supervisors when learning cryptographic verification tools. Figure~\ref{fig:learning-resources-upset} shows that all 16 participants (100\%) used official documentation or tool manuals, making this the most common learning resource. Research papers applying the tools to real protocols were used by 12 participants (75.0\%), while 10 participants (62.5\%) relied on publicly available models, such as GitHub repositories. Peer or supervisor guidance was reported by 9 participants (56.3\%), and 4 participants (25\%) used video lectures or summer-school recordings.

\begin{takeaway}
Participants relied on multiple complementary resources rather than a single learning source (see Figure~\ref{fig:learning-resources-upset}), combining official documentation with research papers, public models, and interpersonal guidance. It indicates that respondents commonly supplemented official documentation with papers, public models, and interpersonal guidance.
\end{takeaway}

\looseness-1
\noindent \textbf{Steepest part of the learning curve. }
When asked about the steepest part of the learning curve, 9 participants (56.3\%) selected debugging non-termination and performance issues. This indicates that the most significant difficulty is not limited to learning the tool's syntax or modeling language, but also involves understanding why proof search becomes slow, fails to terminate, or consumes excessive computational resources. Addressing these problems often requires users to inspect proof states, modify models, introduce helper lemmas, and experiment with proof strategies. Understanding the underlying formal theory, such as applied pi-calculus or multiset rewriting, was identified by 4 participants (25.0\%). Two participants (12.5\%) reported that translating informal protocol descriptions into the tool’s modeling language was the steepest challenge, while only one participant (6.3\%) selected interpreting verification output and attack traces.

\begin{takeaway}
Although learning challenges occurred across modeling, theory, and result interpretation, non-termination and performance debugging emerged as the most frequently reported steepest part of the learning curve. Becoming proficient with a verifier requires learning not only its language and formal foundations, but also how to diagnose and manage the behavior of its proof search.
\end{takeaway}

\noindent \textbf{Programming concepts that helped. }
Participants reported that several programming concepts helped them use cryptographic verification tools more effectively. Step-by-step algorithmic thinking was the most commonly selected concept, reported by 11 of 16 participants (68.8\%). Pattern matching and case analysis were selected by seven participants (43.8\%), while six participants (37.5\%) identified recursion and reasoning through repeated structures as helpful.

\looseness-1
Control flow, higher-order functions, and debugging skills were each selected by four participants (25.0\%). Three participants (18.8\%) reported that immutability was useful, while one participant (6.3\%) selected error handling. Two participants (12.5\%) also mentioned other relevant experience, including knowledge of multiple programming languages and logic programming.

\begin{takeaway}
    Existing programming knowledge can provide useful cognitive scaffolding, but cryptographic verifiers cannot assume a single programming background or rely on conventional programming concepts alone. Learning resources can therefore connect verifier concepts to familiar reasoning patterns while explicitly introducing domain-specific abstractions that do not transfer directly from prior programming experience.
\end{takeaway}

\noindent \textbf{Previous use of general-purpose model checkers: }Participants were also asked whether they had previously used general-purpose model checkers that helped them understand verification concepts. Half of the participants (8, 50.0\%) reported no prior experience with such tools. \emph{SPIN} was the most commonly used model checker, selected by 4(25.0\%), followed by \emph{Maude} with 3(18.8\%) and \emph{UPPAAL} with two participants (12.5\%). \emph{NuSMV} and \emph{PRISM} were each reported by one participant (6.3\%). Two participants (12.5\%) mentioned other formal reasoning tools, including SMT solvers such as \emph{Z3} and \emph{Vampire}, and the \emph{Isabelle theorem prover}. Since participants could report multiple tools, the percentages do not sum to 100\%.

\begin{takeaway}
    Interfaces, documentation, and tutorials cannot assume that new users already understand concepts such as state exploration, formal properties, or counterexample-based reasoning. Cryptographic verification tools must therefore remain learnable for users entering from security or programming backgrounds without previous model-checking experience.
\end{takeaway}

\subsection{Modeling and Verification}
After learning the tools’ basic syntax and workflow, participants still face difficult modeling decisions. Verification depends on whether the formal model accurately captures the protocol, its assumptions, and intended security properties. We therefore examine participants’ confidence in model fidelity, factors limiting that confidence, and strategies used when verification does not proceed as expected.

\noindent \textbf{Confidence in model accuracy. } When asked how confident they were that their formal model accurately reflects the real-world protocol or implementation, participants leaned toward moderate-to-high confidence, though with notable spread. The most common response (7, 43.75\%) was \emph{"Confident"}, while the remaining responses were split evenly (3, 18.75\%) across \emph{"Slightly confident"}, \emph{"Moderately confident"}, and \emph{"Very confident"}. This distribution yields a mean confidence rating of approximately 3.6/5 across the 16 respondents, suggesting that while a majority of participants trust their models to a reasonable degree, a meaningful minority (18.8\%) remain only slightly confident, indicating that confidence in model fidelity is far from uniform across users of these tools.

This pattern is consistent with participants' explanations of what most limits their confidence. The two most frequently cited factors were the lack of a systematic methodology to validate the model against the real protocol (8, 50\%) and difficulty modeling stateful or complex protocol behavior (8, 50\%), followed by having no way to cross-check the model against the actual implementation (7, 43.8\%) and concern that the symbolic model's assumption of perfect cryptography may not reflect real-world implementation weaknesses (6, 37.5\%). Fewer participants (3, 18.75\%) cited uncertainty about the realism of the Dolev-Yao attacker model, and one participant additionally identified limitations inherent to symbolic formal verification.

\begin{takeaway}
    The moderate confidence levels reported in Q9 are driven primarily by structural gaps in the verification workflow, namely the absence of validation methodology and cross-checking mechanisms against real implementations, rather than by fundamental doubts about the symbolic modeling paradigm or the underlying attacker model.
\end{takeaway}

\noindent \textbf{Handling proof failures without a concrete attack trace. } When a proof failed without producing a concrete attack trace, participants' most used strategies were simplifying the model to isolate the source of failure, adding helper or auxiliary lemmas to guide the proof, and re-examining modeling decisions and abstractions for potential unsoundness. These three approaches were selected at identical rates (11, 68.8\%), suggesting that experienced users treat proof failures as a debugging problem to be decomposed methodically rather than a signal to abandon or restart the model. Increasing computational resources or adjusting tool parameters was also fairly common (6, 37.5\%), while consulting published models or peers (3, 18.75\%) and switching tools to cross-validate results (2, 12.5\%) were used less frequently, possibly reflecting the relative isolation in which many users work or the limited availability of comparable published models. Only one participant reported assuming the model was flawed and rewriting it from scratch, and one participant stated:
\\ \\
\emph{``Tweak model to consider a less complex case and see what happens''}. -- P7 

\begin{takeaway}
    Participants addressed proof failures through iterative, model-focused debugging, such as simplifying models, adding auxiliary lemmas, and revisiting assumptions. This places substantial diagnostic burden on users, suggesting a need for better support in identifying problematic model components and explaining unsuccessful proof search.
\end{takeaway}

\cleardoublepage

\looseness-1
\noindent \textbf{Communication during slow or non-terminating verification. } Participants' experiences with tool communication during slow or non-terminating verification runs were predominantly negative. The most frequently selected response (8, 50\%) was that the tool hangs silently with no output or progress indication, followed by the tool producing verbose low-level output that is hard to map back to the participant's model (6, 37.5\%) and the tool showing generic progress output but nothing specific to the model (5, 31.25\%). Fewer participants (3, 18.75\%) reported that the tool at least indicates which rule or lemma is causing the bottleneck, though without offering guidance, and only two participants(12.5\%) described the tool as clearly communicating the source of the problem and how to address it. Overall, most of the selections described feedback as absent, generic, or difficult to map back to the model, whereas only two selections described the tool as clearly communicating both the source of the problem and how to address it. 

\noindent \textbf{Shortcomings in error messaging. } When asked about the most common shortcomings in the tool's error messaging, participants most often pointed to the absence of actionable guidance, with no suggestion or hint on how to fix the issue being the top-selected shortcoming (9, 56.25\%). This was followed by error messages that do not pinpoint the exact location of the problem in the model (8, 50\%) and semantic errors that are silently ignored rather than explicitly reported (7, 43.75\%). A smaller number of participants (5, 31.25\%) noted that error messages assume too much prior knowledge of the underlying formal theory or that errors are too low-level and tied to internal tool mechanics rather than the model itself (3, 18.75\%), while only two participants reported no shortcoming in the tool's error messaging at all.

\begin{takeaway}
    Taken together with the findings on Q12, these results indicate that error and progress messaging represent a substantial usability gap in current cryptographic verification tools: participants are frequently left to diagnose both the location and the cause of a problem largely on their own, whether that problem manifests as a stalled verification run or an unclear error message, with silent failures, either through non-termination or unreported semantic errors, emerging as a recurring theme across both questions.
\end{takeaway}

\subsection{Tamarin specific usability}
\subsubsection{Recovery from false positive counter-example}
Tamarin users may sometimes encounter counterexamples that initially appear to represent valid attacks but are later determined to be false positives. In this study, we use the term false positive to refer to a counterexample that does not correspond to a realistic attack on the intended protocol, but rather arises from modeling errors, under-specified assumptions, abstraction choices, or misunderstandings of the verification output. To examine how users recover from such situations, participants with Tamarin experience were asked whether they had encountered false positive counterexamples and how they typically resolved them.

The responses show that the most common strategy was manually analyzing the counterexample to confirm that it was not a real attack, reported by 10 participants (83.3\%). Eight participants (66.7\%) refined or corrected the protocol model or specification, indicating that false positives often result from mismatches between the intended protocol behavior and its formal representation. Two participants adjusted assumptions about the adversary or environment, while one sought help from peers, supervisors, or online communities. One participant reported another recovery approach.

\begin{takeaway}
    The recovery mainly depends on careful trace interpretation and iterative model refinement. This process can require considerable expertise, suggesting a need for clearer counterexample explanations and better diagnostic support in Tamarin.
\end{takeaway}

\subsubsection{Scalability and Practical Limitations}
To examine the practical usability of Tamarin beyond small examples, participants were asked about the most common limiting factors they encountered when running larger or more realistic case studies. The response options included memory usage, long or non-terminating execution, manual effort required for interactive tracing, tool crashes or instability, not running large models, and other factors.

The responses to this question help capture the scalability-related challenges of using Tamarin in practice. While small protocol models may be easier to construct and verify, larger case studies often introduce additional complexity due to a larger state space, more protocol rules, richer adversary behavior, and more complex security properties. As a result, users may face limitations not only from computational resources, but also from the amount of manual effort required to guide, inspect, or debug the verification process.

Manual effort was the most frequently (8, 66.7\%) reported limitation. This indicates that larger models often require substantial interactive tracing, inspection, and debugging. Memory usage and long or non-terminating execution were each reported by 7 participants (58.3\%), showing that computational resource requirements and verification time are also major scalability concerns. Tool crashes or instability were reported by 2 participants (16.7\%).

\begin{takeaway}
    Tamarin’s scalability is limited by both computational demands and the human effort required to guide and interpret complex verification tasks. Improved automation, resource management, and support for interactive analysis could therefore make larger case studies more practical.
\end{takeaway}

\subsubsection{Reliance on Helper Lemmas.}
Helper lemmas are intermediate properties used to guide Tamarin’s proof search, reduce the search space, and support termination when automated reasoning is insufficient. Among the 12 participants, five (41.7\%) reported that they often needed to write helper lemmas, while three (25.0\%) needed them almost always. Another three participants (25.0\%) required them sometimes, and only one (8.3\%) rarely relied on them. Overall, eight participants (66.7\%) often or almost always needed helper lemmas, while 11 (91.7\%) required them at least sometimes. 

\begin{takeaway}
    These findings indicate that successful verification frequently depends on manually constructed intermediate lemmas, requiring users to understand both the protocol model and Tamarin’s proof-search behavior. Better automated guidance for generating or recommending helper lemmas could therefore reduce manual effort and improve usability.
\end{takeaway}

\subsubsection{Usefulness of dependency graph}
In Tamarin’s interactive mode, the dependency graph visualizes the current proof state by showing rule instances and the dependencies between their premises and conclusions. It helps users examine protocol traces, identify unresolved constraints, and navigate the proof process.

\looseness-1
Among the 12 participants, four (33.3\%) rated the dependency graph as intuitive, while one (8.3\%) considered it very intuitive. Four participants (33.3\%) found it moderately intuitive, indicating that the graph was readable but required effort. Two participants (16.7\%) rated it as unintuitive, and one (8.3\%) considered it very unintuitive. In general, five participants (41.7\%) viewed the visualization positively, while three (25.0\%) found it unintuitive or very unintuitive.

\begin{takeaway}
    Participants' ratings suggest that Tamarin's dependency graph is usable for many users but is not uniformly intuitive. The graph is intended to help users understand and navigate complex proof states; if interpreting the visualization itself requires substantial experience, some of its diagnostic value is lost precisely for users who need additional guidance. Clearer visual organization, navigation, and explanations of proof-state elements could make interactive analysis more accessible.
\end{takeaway}


\subsubsection{Difficulty of modeling protocol features. }
\begin{figure}[t]
    \centering
    \includegraphics[width=\columnwidth]{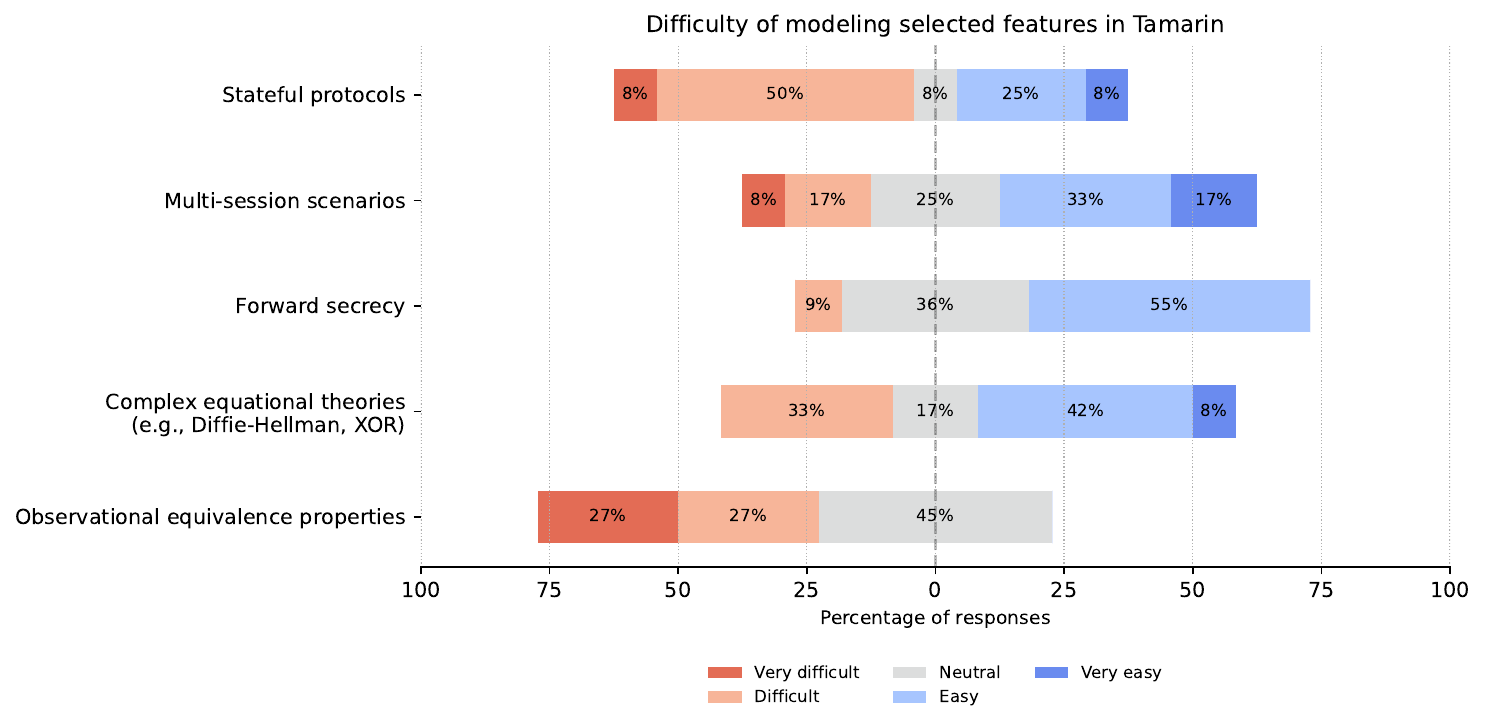}
    \caption{Participants' ratings of the difficulty of modeling
    selected features in Tamarin. Ratings ranged from 1
    (Very difficult) to 5 (Very easy). Numbers inside the bars
    indicate response percentage.}
    \label{fig:tamarin-difficulty}
\end{figure}

Participants rated the difficulty of modeling five protocol features on a five-point scale ranging from \emph{Very difficult} to \emph{Very easy}. As shown in Figure~\ref{fig:tamarin-difficulty}, observational equivalence properties and stateful protocols received the highest proportions of difficult ratings. For observational equivalence, (6, 54\%) of responses were either \emph{Very difficult} or \emph{Difficult}, while the remaining (5, 45\%) were neutral; no participants rated this task as easy. Similarly, (7, 58\%) of the ratings for stateful protocols indicated difficulty, compared with (4, 33\%) indicating easy. 11 participants responded to observational equivalence properties.

By contrast, forward secrecy received the most favorable ratings: (6, 55\%) of participants rated it as easy, (4, 36\%) were neutral, and only 9\% rated it as difficult. Multi-session scenarios also received comparatively positive ratings, with 50\% selecting either \emph{Easy} or \emph{Very easy}, 25\% selecting \emph{Neutral}, and 25\% reporting difficulty. For complex equational theories, including Diffie--Hellman and XOR, half of the responses indicated ease, while 33\% indicated difficulty and 17\% were neutral.

\begin{takeaway}
\looseness-1
Participants reported the greatest difficulty with observational equivalence and stateful protocols. Forward secrecy was viewed as the most approachable feature, while ratings for multi-session scenarios and complex equational theories were comparatively favorable but more varied.
\end{takeaway}

\subsection{ProVerif specific usability}
\label{sec:proverif-usability}
Ratings and representative concerns for Q20--Q24 were derived using the coding procedure described in Section~\ref{sec:data-analysis}: each open-ended justification was assigned an inductively-derived, question-specific code, which was in turn mapped onto a shared five-level adequacy scale (\emph{Fully Adequate}, \emph{Adequate with Minor Limitations}, \emph{Adequate but Requires Significant Improvement}, \emph{Insufficient}, \emph{Not Sure}) to allow comparison across properties. This shared scale underlies the response distributions reported details in Appendix Tables~\ref{tab:q24}--\ref{tab:q20}.

\subsubsection{Adequacy of Observational Equivalence Support (ProVerif)}
Participants expressed mixed views on ProVerif's support for
observational equivalence (i.e., whether an intruder can distinguish between two protocols or scenarios). Appendix Table~\ref{tab:q20} provides the full response distribution and representative concerns from participants' free-text justifications.

When asked how adequate ProVerif's current support for observational equivalence is, responses were fairly evenly distributed across the scale, with a slight lean toward perceiving limitations: four participants rated it insufficient for modern protocols, three rated it adequate but requiring significant improvements, two rated it adequate with only minor limitations, two rated it fully adequate, and two were not sure. Notably, 7 of the 13 respondents (54\%) selected one of the two \emph{Insufficient} or \emph{Requires Significant Improvement} bands, suggesting that a majority of participants who ventured an opinion see room for growth in this area.

The qualitative comments accompanying these ratings point to a few recurring themes rather than a single shared complaint. Several participants raised concerns about the granularity and precision of diff-equivalence, describing it as an over-approximation that is often too coarse or too strong to capture properties like anonymity, unlinkability, or processes that differ in more than simple terms. Others pointed to practical limitations, including termination difficulties and a desire for more specific diagnostic explanations when equivalence checks fail, echoing the broader error-messaging concerns raised in Q13. A smaller number of participants noted domain-specific gaps, such as applicability to IoT and post-quantum computing contexts, or the desire to reason about more than two executions for properties like robust declassification. Interestingly, even among participants who rated the feature as adequate, one noted that usability depends heavily on deep prior knowledge of the tool, reinforcing a theme that surfaces throughout the survey: ProVerif's core theoretical machinery may be sound, but its accessibility and diagnostic transparency remain significant barriers for users.

\subsubsection{Support for Complex Algebraic Properties}
Participants generally viewed ProVerif's support for complex
algebraic properties as limited or dependent on careful modeling. Appendix Table~\ref{tab:q21} provides the detailed responses distribution and representative comments.

Participant ratings of ProVerif's support for complex algebraic properties were mixed but leaned toward identifying limitations: four participants found support limited for algebra-heavy designs, three were not sure, three considered it adequate with careful modeling, and one each rated it fully adequate, prone to false positives, or insufficient and requiring redesign. The most consistent theme across responses, appearing in both the \emph{Limited} and \emph{Requires Redesign} codes, was ProVerif's lack of native support for AC (associative-commutative) operators and Diffie--Hellman-type equations, which participants noted are common in real-world protocol designs and are better supported in tools like Tamarin; several participants explicitly benchmarked ProVerif against Tamarin on this point. Even among participants who considered ProVerif's algebraic support adequate, responses emphasized that this adequacy is conditional: users must carefully construct equational theories and explicitly account for properties such as associativity and commutativity, and one participant noted that termination behavior in the presence of such theories is not always predictable. 
\begin{takeaway}
    Together, these results suggest that algebraic reasoning, particularly involving DH-type and AC equations, remains one of the more technically demanding aspects of modeling in ProVerif, requiring either specialized expertise or acceptance of reduced expressiveness.
\end{takeaway}

\subsubsection{Support for State-Dependent Protocols}
Responses tended to identify limitations in ProVerif's support for state-dependent protocols, particularly for complex stateful designs. Appendix Table~\ref{tab:q22} reports the full distribution and representative concerns. Responses to Q22 skewed toward viewing ProVerif's support for state-dependent protocols as inadequate: four participants rated it insufficient for complex stateful designs, three rated it adequate but requiring significant improvements, and only two rated it fully adequate. Combined, 7 of 13 respondents (54\%) selected one of the two lowest-rated bands. A recurring theme across responses was the reliance on tables as a workaround for modeling state, which multiple participants described as usable but unsatisfactory, requiring substantial manual effort, and of uncertain scalability to larger protocols. Several participants pointed to concrete domains where this limitation is especially acute, such as protocols involving TPMs or IoT devices, and one noted having to significantly deviate from the original protocol specification simply to achieve termination. Extensions such as GSVerif were mentioned as a partial mitigation, and debugging stateful models was separately flagged as a source of difficulty even when a workable encoding was found. Several participants also expressed hope that native support for mutable state would appear in future tool versions. 
\begin{takeaway}
    Respondents reported burdens when encoding richer state evolution using ProVerif's available mechanisms.
\end{takeaway}

\subsubsection{Adequacy of Support for Perfect Forward Secrecy}

\looseness-1
Appendix Table~\ref{tab:q23} summarizes participant ratings of ProVerif's support for verifying Perfect Forward Secrecy (PFS).
For Perfect Forward Secrecy, the dominant response was uncertainty: seven of 13 participants (54\%) selected \emph{Not Sure}, considerably more than for any other adequacy question in the survey. Among participants who did express an opinion, responses were generally positive, with three rating ProVerif's support fully adequate and two rating it adequate with only minor limitations. The concerns raised by these participants were relatively narrow in scope, centering on the need to carefully model the notion of a compromised ``past'' state and a suggestion that testing a greater number of sessions could strengthen confidence in verification results. Only one participant rated support as requiring significant improvement, noting that modeling PFS still requires ad hoc tricks. The high proportion of \emph{Not Sure} responses suggests that, relative to other properties surveyed, participants have comparatively less direct experience modeling Perfect Forward Secrecy in ProVerif, which limits how strongly conclusions about its adequacy can be drawn from this sample.
\subsubsection{Support for Post-Compromise Security}
\looseness-1
Participant ratings of ProVerif's support for post-compromise security, which requires reasoning about state evolution are shown in Appendix Table~\ref{tab:q24}. As with Perfect Forward Secrecy, uncertainty dominated responses regarding post-compromise security, with seven of 13 participants (54\%) selecting \emph{Not Sure}. Among the remaining participants, opinions were split evenly between viewing ProVerif's support as workable and viewing it as insufficient: two participants rated it adequate with careful modeling, citing existing models for protocols such as Signal as evidence of feasibility while explicitly noting that such models require substantial manual effort to construct, and two rated it insufficient for modern ratcheting protocols, with one hoping that related overfitting issues would be addressed in future updates despite acknowledging this as a hard problem. Only one participant rated support as fully adequate, and one rated it adequate but limited specifically for dynamic recovery scenarios, noting that this remains difficult to model in a straightforward way. The high rate of uncertainty for Q24 suggests that many participants had limited direct experience modeling post-compromise security in ProVerif. Participants who did express an opinion frequently emphasized the modeling effort required. These responses should therefore be interpreted as usability judgments rather than bounds on ProVerif’s technical capabilities. Recent work~\cite{cheval2026automated} has demonstrated a highly automated ProVerif-based analysis of Signal’s Double Ratchet that reasons about key compromise and proves post-compromise security in the analyzed model, showing that sophisticated PCS verification is possible while potentially requiring specialized modeling expertise.

\subsubsection{Cross-Cutting Themes in ProVerif Adequacy}
Beyond the property-specific concerns above, our axial coding pass across Q20--Q24 identified several themes that recurred across multiple properties rather than being specific to any single one. \emph{Manual effort} was repeatedly cited as the qualifier that separates a nominally adequate feature from a practically usable one, appearing in participants' justifications for complex algebraic properties (Q21), state-dependent protocols (Q22), and post-compromise security (Q24), and echoing the workaround-heavy debugging strategies reported earlier in Q11. \emph{Termination behavior} surfaced as an independent source of uncertainty across observational equivalence, algebraic properties, and stateful modeling (Q20--Q22), reinforcing the non-termination concerns raised throughout the broader modeling and verification workflow (Q12). Participants also explicitly \emph{benchmarked ProVerif against Tamarin}, particularly regarding native support for Diffie--Hellman-type and AC-operator reasoning (Q21), and raised \emph{domain-specific applicability} concerns for IoT, TPM, and post-quantum settings (Q20, Q22). Several participants expressed \emph{hope for native support} of currently unsupported features, such as mutable global state, in future tool versions (Q22, Q24), while others pointed to \emph{existing published models} (e.g., for Signal) as evidence that even demanding properties can be verified given sufficient effort (Q24). 
\begin{takeaway}
   Participants' concerns about ProVerif extend beyond any single security property to shared structural limitations, namely termination predictability, native algebraic and state support, and the manual effort required to work around their absence. 
\end{takeaway}

\subsubsection{Desired Features and Improvements for ProVerif} When asked which features or improvements would most increase their productivity with ProVerif, participants most frequently selected a model validation mechanism to check whether the formal model matches the intended protocol (n = 10), followed by an interactive debugging mode allowing step-by-step inspection of the proof state (n = 8) and a graphical interface for model visualization and attack trace exploration (n = 6). Improved documentation with realistic, well-annotated protocol examples was also fairly commonly selected (n = 5), while integration with standard development tools such as IDE plugins or CI/CD pipelines (n = 3) and automated suggestions for fixing common modeling mistakes (n = 3) were selected less frequently. Two participants used the open-ended option to specifically request automated generation of helper lemmas.

This ranking of desired improvements aligns closely with limitations identified elsewhere in the survey. The top-ranked request, a model validation mechanism, directly mirrors the leading concerns raised in Q10, where the lack of a systematic methodology to validate the model against the real protocol and the inability to cross-check the model against the actual implementation were among the most commonly cited limits on confidence. Similarly, the demand for an interactive debugging mode and a graphical interface for trace exploration reflects the debugging-heavy troubleshooting strategies reported in Q11 and the recurring complaints in Q12 and Q13 about silent failures and unhelpful error messages. The relatively lower demand for IDE or CI/CD integration and automated mistake-fixing suggests that participants' priorities center less on tooling convenience and more on fundamental gaps in validation, transparency, and debuggability.
\begin{takeaway}
    Users are less concerned with ProVerif fitting into modern development workflows than with being able to trust and understand the models and proofs it produces.
\end{takeaway}

\section{RQ1: Usability Barriers in Cryptographic Protocol Verification Tools}
\looseness-1
\noindent\textbf{Barriers across the verification workflow.} Usability barriers occur throughout the verification workflow rather than at a single stage. Learning specialized syntax and formal theory is challenging, but the more persistent difficulties arise during model construction, proof management, and interpreting outcomes: translating informal protocols into suitable symbolic abstractions, judging whether those abstractions are accurate, and managing proof searches that may stall or fail to terminate.

\looseness-1
\noindent\textbf{Model fidelity and validation.} A central barrier is confidence in model fidelity. Even confident participants reported limited support for systematically validating a formal model against the original protocol or implementation, a gap between verifying a representation and establishing that it faithfully captures the real system. Difficulty representing stateful and complex behavior widens this uncertainty, particularly when the verifier requires workarounds that diverge from the original specification.

\looseness-1
\noindent\textbf{Proof debugging and diagnostic feedback.} Proof failures impose a substantial manual debugging burden. Users typically respond by simplifying models, revisiting assumptions, and introducing helper lemmas, work that requires understanding both the protocol and the verifier's proof-search behavior. Limited progress information and non-actionable error messages compound this: silent hangs, low-level internal output, or unlocalized errors leave users to diagnose problems largely through experimentation and prior expertise.

\looseness-1
\noindent\textbf{Tool-specific modeling and proof challenges.} Tool-specific findings reinforce these broader barriers. Tamarin users frequently depend on helper lemmas, interactive proof guidance, and manual counterexample analysis; its dependency graph can support expert analysis but may require substantial experience to interpret, and modeling stateful protocols or observational equivalence remains especially demanding. ProVerif participants identified similar difficulties with state-dependent protocols, complex algebraic properties, and observational equivalence, properties that workarounds can represent only at the cost of significant manual effort and reduced fidelity to the original protocol.

\looseness-1
\noindent\textbf{Gap between protocol reasoning and verifier behavior.} The main usability challenge is therefore not learning to operate a verifier, but managing the gap between the user's protocol-level reasoning and the tool's formal representation, proof procedures, and diagnostic output. Validating models, guiding proofs, diagnosing non-termination, and interpreting results remain largely the user's responsibility, so effective use depends on substantial formal-methods expertise, which may limit accessibility and wider adoption.

\section{RQ2: Perceived Transfer of Programming and Model-Checking Knowledge}

\looseness-1
\noindent\textbf{Transferable programming strategies.}
Prior programming experience transfers mainly through general problem-solving strategies. Participants drew on step-by-step algorithmic reasoning, pattern matching, case analysis, recursion, and debugging when constructing and analyzing models, supporting decomposition, reasoning about alternative execution paths, and diagnosis. However, no single concept was universally essential, and programming experience does not fully prepare users for symbolic reasoning, unbounded sessions, attacker knowledge, equational theories, or interpreting formal security guarantees, so users must adapt familiar reasoning to a substantially different setting.

\looseness-1
\noindent\textbf{Model checking as a conceptual bridge.}
Prior experience with general-purpose model checkers may provide a more direct bridge: familiarity with state exploration, property specification, and counterexamples helps users understand the broader purpose and behavior of cryptographic verifiers. However, half of participants had no such experience, so this background is helpful but not a prerequisite, and verifiers cannot assume new users already understand these concepts.

\looseness-1
\noindent\textbf{Limits of knowledge transfer.}
Cryptographic verification introduces domain-specific concepts not typically found in conventional programming or model checking, including symbolic adversary models, cryptographic term algebras, protocol-specific security properties, and formalisms such as multiset rewriting and the applied pi calculus. Prior knowledge may clarify the shape of the verification task, but does not eliminate the need to learn these specialized theories and conventions.

\looseness-1
\noindent\textbf{Implications for onboarding.}
Programming and model-checking knowledge together provide a helpful but incomplete foundation: programming contributes transferable reasoning and debugging strategies, while model checking provides familiarity with verification concepts and workflows. Neither fully addresses the specialized abstractions and proof-management demands of cryptographic protocol analysis, so onboarding materials should explicitly connect familiar concepts to their verification counterparts while supporting users without prior formal-methods experience.

\section{RQ3: Usability, Design, and Functional Improvements}
Improving these tools requires more than simplifying syntax or modernizing interfaces. Participants prioritized reducing uncertainty during model construction, making proof behavior transparent, and strengthening support for diagnosing unsuccessful verification. Across both tools, the most important needs concerned model validation, actionable diagnostics, counterexample explanation, interactive debugging, and assistance with complex proof tasks. An overview appears in Appendix Table~\ref{tab:requested-improvements}.

\looseness-1
\noindent \textbf{Model construction and validation support.}
A model-validation mechanism was the most requested ProVerif improvement, helping users compare formal models with protocol descriptions, identify omitted assumptions, and link model elements to implementation behavior. It would not guarantee correctness but could reduce mismatches between the verified model and the intended protocol.

\looseness-1
\noindent \textbf{Error feedback and counterexample explanation.}
For Tamarin, error feedback and counterexample explanation were the top requests (eight responses each). Participants wanted diagnostics that identify the relevant rule, lemma, query, or construct and explain possible causes and fixes, with counterexamples summarized at the protocol level so genuine attacks can be distinguished from modeling errors or abstraction artifacts.

\looseness-1
\noindent \textbf{Interactive debugging and visualization.}
Interactive debugging and visualization could reduce the manual effort needed to understand verification behavior. Eight ProVerif participants wanted step-by-step proof-state inspection, and six wanted a graphical interface for model and trace exploration. Such interfaces should connect internal proof operations to the user's original protocol representation, letting users trace attacker knowledge, identify unresolved branches, and locate where search becomes unmanageable. For Tamarin, this means clearer dependency-graph layouts, better navigation, and stronger interactive/proof-oracle support for case distinctions; for ProVerif, an interactive mode for examining failed queries and diagnosing non-termination that is currently hard to infer from static output.

\looseness-1
\noindent \textbf{Automation and proof guidance.}
Tamarin users frequently depend on manually written helper lemmas and case distinctions, and two ProVerif respondents requested automated lemma generation. Future tools could analyze recurring proof patterns to recommend intermediate lemmas, case distinctions, model simplifications, or alternative strategies, and detect modeling mistakes or abstractions that produce false counterexamples or non-termination. AI-assisted functionality could also synthesize counterexamples, summarize proof states, and explain unresolved branches, but such recommendations should remain transparent and reviewable, supporting rather than obscuring the formal basis of the result.

\looseness-1
\noindent \textbf{Documentation and tooling support.}
Documentation and tooling support were each selected by five Tamarin participants, and five ProVerif respondents wanted improved, realistic, well-annotated examples. Traditional reference manuals appear insufficient alone; documentation should explain modeling choices, rejected alternatives, common mistakes, and debugging steps, ideally with comparative examples across tools for non-termination, false counterexamples, helper lemmas, and stateful or equivalence properties. Tooling support could include syntax-aware editors, inline diagnostics, navigation between models and counterexample traces, and state-transition visualization. IDE and CI/CD integration was a lower priority (three respondents), suggesting adoption is currently constrained more by validation, interpretability, and debuggability than by workflow convenience; integration is likely to matter more once users can trust the verification process.

\looseness-1
\noindent \textbf{Functional support for complex protocols and properties.}
Some barriers require changes beyond interfaces or documentation. For Tamarin, participants wanted stronger support for stateful protocols, observational equivalence, proof scalability, and interactive analysis; for ProVerif, improvements to mutable state, ordered transitions, complex algebraic theories, observational equivalence, and properties involving state evolution such as post-compromise security. Addressing these may require changes to the underlying frameworks, reducing reliance on workarounds that increase manual effort and distance the model from the original protocol.
\section{Acknowledgments}
We thank all survey participants for generously contributing their time and sharing their experiences and perspectives. We are particularly grateful to Qifan Zhang of Palo Alto Networks and Zilin Shen of Purdue University for participating in the pilot survey and providing valuable feedback that informed the refinement and finalization of the questionnaire. Their input helped improve the clarity, relevance, and overall quality of the survey instrument.

\section{Conclusion}

\looseness-1
We presented an exploratory survey of 16 experienced users of cryptographic protocol verification tools, including Tamarin and ProVerif. Participants reported challenges in validating whether formal models faithfully represent intended protocols, diagnosing slow or non-terminating verification, and interpreting unsuccessful proofs and tool feedback. Prior programming and model-checking experience provided useful reasoning strategies but did not replace the domain-specific knowledge required for symbolic protocol verification. Across tools, participants emphasized the need for better model-validation support, actionable diagnostics, clearer explanations of verification outcomes, interactive debugging, and greater automation of recurring proof tasks.

Our findings should be interpreted in light of the study’s small, self-selected, and predominantly experienced sample and its reliance on self-reported rather than observed tool use. Overall, improving cryptographic verifiers requires reducing the gap between users’ protocol-level reasoning and the tools’ formal representations and proof procedures. Future work should evaluate these needs through observational studies and concrete mechanisms for model validation, proof guidance, and explanation.

\appendix
\section*{Ethical Considerations}
\looseness-1
\textbf{Human Subjects and Informed Consent.} This study involved an online questionnaire with researchers, graduate students, and industry practitioners who had hands-on experience with cryptographic protocol verification tools. The study was reviewed and approved by the relevant Institutional Review Board (IRB). Before beginning the survey, participants were presented with a consent form describing the purpose and scope of the study, the survey procedure, potential risks and benefits, and other information relevant to participation. Participants could proceed with the survey only after explicitly providing consent by selecting the “I agree” option; participants who selected “I disagree” were not permitted to continue.

\looseness-1
\noindent \textbf{Participant Privacy and Confidentiality.} Because cryptographic protocol verification is a relatively small and specialized research community, combinations of professional role, experience, tool usage, research activities, or qualitative responses could potentially make individual participants identifiable even after direct identifiers are removed. This creates potential professional or reputational risks, particularly when participants describe difficulties they encountered, limitations of particular tools, or their own verification practices. To mitigate these risks, we refer to participants using anonymous identifiers, remove potentially identifying details from quoted responses, and avoid reporting project-specific information that could reveal participant identities. We also avoid unnecessarily linking qualitative responses with detailed participant characteristics when such combinations could increase re-identification risk.

\noindent \textbf{Data Protection and Reporting.} Survey responses were collected through Qualtrics and stored securely for research purposes. We report quantitative results primarily in aggregate and use de-identified excerpts when presenting qualitative responses. Because the participant population is specialized and some combinations of demographic characteristics and tool experience may be distinctive, we take particular care when reporting participant-level information. Any materials released alongside the paper will be reviewed to ensure that they do not expose direct identifiers or combinations of information that could reasonably facilitate re-identification.

\looseness-1
\noindent \textbf{Affected Stakeholders.} We identify the following primary stakeholder groups that may be affected by our work: (1) \textit{Study participants: }are the primary stakeholders because they provided potentially sensitive descriptions of their expertise, workflows, difficulties, and opinions about particular verification tools. Given the relatively small and specialized cryptographic-verification community, we considered re-identification and associated professional or reputational harm to be the primary risks and therefore limited the disclosure and linkage of participant-specific information; (2) \textit{Tool developers and maintainers:} are also affected because our findings identify perceived limitations and recommend changes to existing tools. We therefore distinguish participants' reported experiences from objective technical deficiencies and avoid attributing usability problems or design trade-offs to individual developers; (3) \textit{Current and prospective users:} may benefit from improved diagnostics, model-validation support, documentation, visualization, and proof guidance, while inaccurate or overly broad interpretations of our findings could give users misleading expectations about the capabilities or limitations of particular tools; (4) \textit{The broader security and formal-methods research community:} may use these findings to inform future tool design and human-centered research. Because our evidence is based on a small, specialized, self-selected sample, we avoid claiming that the reported experiences are representative of all users of cryptographic verification tools.

\looseness-1
\noindent \textbf{Responsible Characterization of Tools and Findings.} Our study examines users’ experiences with cryptographic verification tools rather than evaluating the competence or intentions of their developers. Some challenges reported by participants, including non-termination, limitations in modeling particular protocol features, or difficulties associated with specific abstractions, may arise from fundamental properties or design trade-offs of the underlying verification approaches rather than straightforward software defects. We therefore distinguish between participants’ reported experiences, our interpretation of those experiences, and objective claims about a tool’s technical capabilities. Similarly, difficulty using a verification tool should not be interpreted as evidence of insufficient expertise on the part of an individual participant. Our goal is to identify opportunities for improving usability and supporting users rather than assigning responsibility for the difficulties reported.

\noindent \textbf{Respect for Law and Public Interest.} The study consisted of a questionnaire about participants’ experiences with cryptographic protocol verification tools and did not involve probing deployed systems, exploiting vulnerabilities, or interacting with third-party systems without authorization. The survey focused on participants’ learning experiences, modeling and verification workflows, difficulties encountered during proof construction and debugging, and desired improvements to existing tools. We avoid reporting identifying project details or other information that could create unnecessary risks to participants or third parties. The expected public benefit of this research is a better understanding of the usability barriers surrounding cryptographic protocol verification and the identification of design priorities that may help researchers and practitioners construct, debug, interpret, and validate formal protocol models more effectively.

\noindent \textbf{Responsible Interpretation and Generalization.} Participants were recruited from a specialized community through academic publications, open-source contributions, community outreach, professional contacts, referrals, and snowball sampling. Participation was voluntary, and the recruitment strategy may introduce self-selection bias. The study also relies on participants’ self-reported experiences rather than direct observation of their verification activities. We therefore interpret the findings as evidence about the experiences and practices reported by our participants and do not claim that the results are statistically representative of all Tamarin, ProVerif, or cryptographic-verification users.

\section*{Open Science}
To support transparency, reproducibility, and evaluation of our study, we make the study materials and analysis artifacts publicly available. During the anonymous review period, these materials are available through an anonymized repository at \url{https://anonymous.4open.science/r/protocol-verifier-usability-artifact-6625/}.
The research artifacts associated with this study are:
\begin{itemize}
    \item Survey questionnaire
    \item Consent form
    \item Codebook
\end{itemize}
\textbf{Materials Shared:} We include the complete questionnaire used in the study, including the consent information. We also share the lightweight \textit{codebook} used to analyze the explanatory free-text responses, with codes, definitions, and example quotes that we coded for each code.

\noindent \textbf{Materials Not Shared:} We do not publicly release the raw Qualtrics response data. Although the survey did not intentionally collect direct personally identifiable information, the cryptographic-verification community is relatively small and specialized, and combinations of response content and survey metadata could increase the risk of participant re-identification. To minimize this risk and preserve participant anonymity, we share only de-identified, processed, and aggregated data necessary to support the analyses reported in this paper, rather than the original Qualtrics export.

\bibliographystyle{plainurl}
\bibliography{references}

\section{Survey Instruments}
\label{app:survey}

{\footnotesize
This appendix presents the full text of the survey administered to
participants with experience using cryptographic protocol
verification tools (Tamarin Prover, ProVerif, and other). The
survey was organized into five blocks: Demographic, Learnability,
Modeling \& Verification, Tamarin, and ProVerif. Unless otherwise
noted, single-select items are marked \emph{(select one)} and
multi-select items are marked \emph{(select all that apply)} or checkbox.

\subsection{Block: Demographic}

\begin{itemize}

\surveyitem Which of the following verification tools have you used?
\emph{(select all that apply)}
\begin{todolist}
    \item Tamarin
    \item ProVerif
    \item DeepSec
    \item Other (please specify)
\end{todolist}

\surveyitem How long have you been using formal verification tools for
security protocols?
\begin{enumerate}
    \item Less than 1 year
    \item 1--3 years
    \item 4--6 years
    \item More than 6 years
\end{enumerate}

\surveyitem What is your current role or background?
\begin{enumerate}
    \item Undergraduate student (studying relevant field)
    \item Graduate student (Master's or PhD)
    \item Academic researcher or faculty
    \item Industry professional (practitioner using formal methods)
    \item Other (please specify)
\end{enumerate}

\surveyitem How would you rate your expertise in formal methods /
cryptographic protocol verification?
\begin{enumerate}
    \item Novice: new to formal verification (just learning or only
    basic knowledge)
    \item Intermediate: comfortable with core concepts and have used
    verification tools on some projects
    \item Advanced: very proficient with formal verification;
    extensive experience using these tools
    \item Deep expertise (e.g., you contribute to tool development
    or advanced research in this area)
\end{enumerate}

\end{itemize}

\subsection{Block: Learnability}

\begin{itemize}

\surveyitem What resources did you rely on when learning the tool?
\emph{(select all that apply)}
\begin{todolist}
    \item Official documentation and tool manual
    \item Research papers applying the tool to real protocols
    \item Video lectures or summer school recordings
    \item Publicly available models from other researchers (e.g.,
    GitHub)
    \item Peer or supervisor guidance
    \item Other (please specify)
\end{todolist}

\surveyitem What was the steepest part of the learning curve?
\emph{(select one)}
\begin{enumerate}
    \item Translating informal protocol descriptions into the
    tool's modeling language
    \item Understanding the underlying formal theory (e.g., applied
    pi-calculus, multiset rewriting)
    \item Interpreting verification output and attack traces
    \item Debugging non-termination and performance issues
    \item Knowing what the proof actually guarantees in practice
    \item Other (please specify)
\end{enumerate}

\surveyitem Which programming concepts from your prior experience helped
you be more effective when learning or using this cryptographic
verification tool? \emph{(select all that apply)}
\begin{todolist}
    \item Step-by-step algorithmic thinking
    \item Control flow (if/else, loops)
    \item Debugging skills (tracing execution, logging, breakpoints)
    \item Error handling (exceptions, return codes, defensive checks)
    \item Higher-order functions
    \item Immutability (avoiding unintended state changes)
    \item Pattern matching / case analysis
    \item Recursion (reasoning via repeated structure)
    \item Other (please specify)
\end{todolist}

\surveyitem Have you previously used any general-purpose model checker
that helped you learn or use this cryptographic verification tool?
\begin{todolist}
    \item SPIN
    \item NuSMV
    \item UPPAAL
    \item TLA+ (TLC)
    \item PRISM
    \item Maude
    \item Other (please specify)
    \item None
\end{todolist}

\end{itemize}

\subsection{Block: Modeling \& Verification}

\begin{itemize}

\surveyitem How confident are you that your formal model accurately
reflects the real-world protocol you intend to verify?
\begin{enumerate}
    \item 1 -- Not confident at all
    \item 2 -- Slightly confident
    \item 3 -- Moderately confident
    \item 4 -- Confident
    \item 5 -- Very confident
\end{enumerate}

\surveyitem What most limits your confidence in the accuracy of your
model? \emph{(select all that apply)}
\begin{todolist}
    \item The symbolic model assumes perfect cryptography, which may
    not reflect actual weaknesses in real-world implementations
    \item Difficulty modeling stateful or complex protocol behavior
    \item Uncertainty about whether the Dolev--Yao attacker model is
    realistic enough
    \item Lack of a systematic methodology to validate the model
    against the real protocol
    \item No way to cross-check the model against the actual
    implementation
    \item Other (please specify)
\end{todolist}

\surveyitem When a proof fails without a concrete attack trace, what is
your typical approach? \emph{(select all that apply)}
\begin{todolist}
    \item Simplify the model to isolate which part is causing the
    failure
    \item Add helper lemmas or auxiliary lemmas to guide the proof
    (e.g., in Tamarin)
    \item Re-examine modeling decisions and abstractions for
    potential unsoundness
    \item Consult related published models or ask peers for
    guidance
    \item Switch to a different tool to cross-validate the result
    \item Assume the model is flawed and rewrite it from scratch
    \item Increase computational resources or adjust tool parameters
    and re-run
    \item Other (please specify)
\end{todolist}

\surveyitem Which best describes the tool's communication when
verification is slow or non-terminating? \emph{(select all that
apply)}
\begin{todolist}
    \item The tool hangs silently with no output or progress
    indication
    \item The tool shows generic progress output but nothing
    specific to my model
    \item The tool produces verbose low-level output that is hard to
    map back to my model
    \item The tool indicates which rule or lemma is causing the
    bottleneck but offers no guidance
    \item The tool clearly communicates the source of the problem
    and how to address it
\end{todolist}

\surveyitem What best describes the most common shortcoming in the tool's
error messaging? \emph{(select all that apply)}
\begin{todolist}
    \item Error messages do not pinpoint the exact location of the
    problem in the model
    \item Errors are too low-level and tied to internal tool
    mechanics rather than the model itself
    \item Semantic errors are silently ignored rather than
    explicitly reported
    \item No suggestion or hint is provided on how to fix the issue
    \item Error messages assume too much prior knowledge of the
    underlying formal theory
    \item No shortcoming in the tool's error messaging
    \item Other (please specify)
\end{todolist}

\end{itemize}

\subsection{Block: Tamarin}

\begin{itemize}

\surveyitem Have you ever encountered a counterexample in Tamarin that you
later determined to be a false positive? If so, how did you
typically recover from or resolve it?
\begin{todolist}
    \item Refined or corrected the protocol model/specification
    \item Manually analyzed the counterexample to confirm it was not
    a real attack
    \item Consulted documentation, examples, or academic papers
    \item Asked for help from peers, supervisors, or online
    communities
    \item Adjusted assumptions about the adversary or environment
    \item Other (please specify)
\end{todolist}

\surveyitem When running larger/realistic case studies in Tamarin, what is
your most common limiting factor(s)?
\begin{todolist}
    \item Memory (RAM)
    \item Time (runs too long / non-terminating)
    \item Manual effort (requires deep interactive tracing)
    \item Tool crashes / instability
    \item I don't run large models
    \item Other (please specify)
\end{todolist}

\surveyitem How often do you need to write helper lemmas to guide
Tamarin's proof search toward termination?
\begin{enumerate}
    \item Never
    \item Rarely
    \item Sometimes
    \item Often
    \item Almost always
\end{enumerate}

\surveyitem How intuitive is the visualization of the dependency graph in
Tamarin's interactive mode?
\begin{enumerate}
    \item 1 -- Very unintuitive: the graph is difficult to read and
    interpret
    \item 2 -- Unintuitive: requires significant experience to
    navigate
    \item 3 -- Moderate: readable with effort but could be clearer
    \item 4 -- Intuitive: the graph is generally easy to read and
    navigate
    \item 5 -- Very intuitive: the graph clearly communicates the
    proof state
\end{enumerate}

\surveyitem How difficult is it to model the following in Tamarin? Rate
each on a scale from 1 (Very Difficult) to 5 (Very Easy).
\begin{todolist}
    \item Stateful protocols
    \item Multi-session scenarios
    \item Forward secrecy
    \item Complex equational theories (e.g., Diffie--Hellman, XOR)
    \item Observational equivalence properties
\end{todolist}

\surveyitem What is the important usability improvement you would like to see in Tamarin Prover?
\begin{todolist}
    \item Modeling support
    \item Error feedback
    \item Counterexample explanation
    \item Documentation
    \item Tooling support
    \item Other (please specify)
\end{todolist}

\end{itemize}

\subsection{Block: ProVerif}

\begin{itemize}

\surveyitem How adequate is ProVerif's current support for observational
equivalence (two protocol executions are indistinguishable to any
attacker) in analyzing modern privacy-preserving protocols? Please
explain your reasoning for your choice.
\begin{enumerate}
    \item Fully adequate
    \item Adequate with minor limitations
    \item Adequate but requires significant improvements
    \item Insufficient for modern protocols
    \item Not sure
\end{enumerate}

\surveyitem How adequate is ProVerif's current support for reasoning about
complex algebraic properties (e.g., Diffie--Hellman, XOR, group
equations)? Please explain your reasoning for your choice.
\begin{enumerate}
    \item Fully adequate for modern protocols
    \item Adequate with careful modeling
    \item Adequate but prone to false positives
    \item Limited for algebra-heavy designs
    \item Insufficient and requires redesign
    \item Not sure
\end{enumerate}

\surveyitem Considering that ProVerif does not natively support mutable
global state or ordered transitions, do you believe its current
framework is fundamentally adequate for verifying stateful protocols
(e.g., TPMs, IoT devices, e-passports)? Please explain your reasoning
for your choice.
\begin{enumerate}
    \item Fully adequate
    \item Adequate with minor limitations
    \item Adequate but requires significant improvements
    \item Insufficient for complex stateful designs
    \item Not sure
\end{enumerate}

\surveyitem In your opinion, how adequate is ProVerif's support for
verifying Perfect Forward Secrecy (compromise of long-term keys does
not break past sessions)? Please explain your reasoning for your
choice.
\begin{enumerate}
    \item Fully adequate
    \item Adequate with minor limitations
    \item Adequate but requires significant improvements
    \item Insufficient for modern protocols
    \item Not sure
\end{enumerate}

\surveyitem Considering that post-compromise security requires reasoning
about state evolution, key updates, and temporal ordering (e.g.,
compromise occurs, then secure recovery), do you believe ProVerif's
current Horn-clause-based over-approximation is fundamentally
adequate for this? Please explain your reasoning for your choice.
\begin{enumerate}
    \item Fully adequate
    \item Adequate with careful modeling
    \item Adequate but limited for dynamic recovery
    \item Insufficient for modern ratcheting protocols
    \item Not sure
\end{enumerate}

\surveyitem Which features or improvements would most increase your
productivity with ProVerif?
\begin{todolist}
    \item An interactive debugging mode allowing step-by-step
    inspection of the proof state
    \item A graphical interface for model visualization and attack
    trace exploration
    \item Automated suggestions for fixing common modeling mistakes
    \item Improved documentation with realistic, well-annotated
    protocol examples
    \item Integration with standard development tools (e.g., IDE
    plugins, CI/CD pipelines)
    \item A model validation mechanism to check whether the formal
    model matches the intended protocol
    \item Other (please specify)
\end{todolist}

\end{itemize}
}

\begin{figure}[t]
    \centering
    \includegraphics[width=0.9\columnwidth]{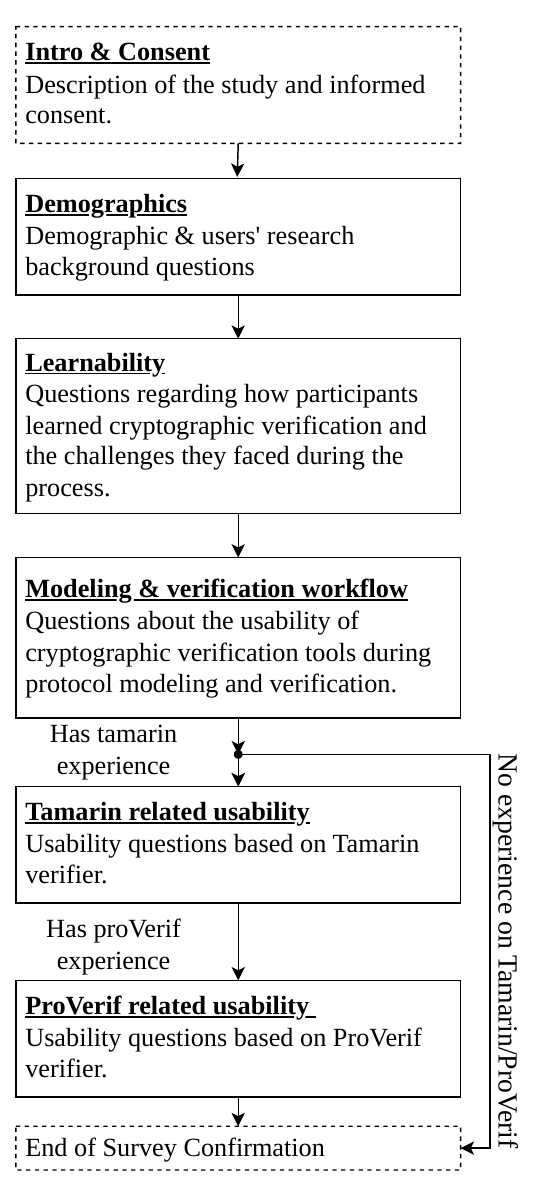}
    \caption{Overview of the survey flow and conditional tool-specific
    branches.}
    \label{fig:study-procedure}
\end{figure}

\section{Figures}
\label{app:participant-demographics}

\begin{table*}[t]
\centering
\caption{Participant ratings of ProVerif's support for post-compromise security.}
\label{tab:q24}
\small
\renewcommand{\arraystretch}{1.3}
\begin{tabular}{@{}p{5cm} p{10cm}@{}}
\toprule
\textbf{Response} & \textbf{Representative concerns raised} \\
\midrule
Not sure (7) & --- \\
\addlinespace
Adequate with careful modeling (2) &
\begin{tabular}[t]{@{}p{10cm}@{}}
$\bullet$ Models exist for protocols such as Signal, but require substantial manual effort \\
$\bullet$ Reference to a recent Signal ProVerif model (S\&P '26)
\end{tabular} \\
\addlinespace
Insufficient for modern ratcheting protocols (2) &
\begin{tabular}[t]{@{}p{10cm}@{}}
$\bullet$ Current support described as inadequate \\
$\bullet$ Hope that issues related to overfitting will be addressed in future updates, despite this being a challenging problem
\end{tabular} \\
\addlinespace
Fully adequate (1) & --- \\
\addlinespace
Adequate but limited for dynamic recovery (1) &
\begin{tabular}[t]{@{}p{10cm}@{}}
$\bullet$ Not straightforward to model
\end{tabular} \\
\bottomrule
\end{tabular}
\end{table*}

\begin{table*}[t]
\centering
\caption{Participant ratings of ProVerif's support for verifying Perfect Forward Secrecy.}
\label{tab:q23}
\small
\renewcommand{\arraystretch}{1.3}
\begin{tabular}{@{}p{5cm} p{10cm}@{}}
\toprule
\textbf{Response} & \textbf{Representative concerns raised} \\
\midrule
Not sure (7) & --- \\
\addlinespace
Fully adequate (3) & --- \\
\addlinespace
Adequate with minor limitations (2) &
\begin{tabular}[t]{@{}p{10cm}@{}}
$\bullet$ Care is needed in how ``past'' states are modeled \\
$\bullet$ Increasing the number of sessions tested may improve confidence in results
\end{tabular} \\
\addlinespace
Adequate but requires significant improvements (1) &
\begin{tabular}[t]{@{}p{10cm}@{}}
$\bullet$ Requires tricks to model correctly
\end{tabular} \\
\bottomrule
\end{tabular}
\end{table*}

\begin{table*}[t]
\centering
\caption{Participant ratings of ProVerif's support for state-dependent protocols.}
\label{tab:q22}
\small
\renewcommand{\arraystretch}{1.3}
\begin{tabular}{@{}p{5cm} p{10cm}@{}}
\toprule
\textbf{Response} & \textbf{Representative concerns raised} \\
\midrule
Insufficient for complex stateful designs (4) &
\begin{tabular}[t]{@{}p{10cm}@{}}
$\bullet$ Modeling attempts required deviating significantly from the protocol specification just to get ProVerif to terminate \\
$\bullet$ Particularly limiting for stateful devices such as TPMs and IoT devices \\
$\bullet$ Tables can be used to emulate a notion of state, but this is unsatisfactory and likely does not scale to large protocols \\
$\bullet$ Hope for future updates to natively support mutable global state and ordered transitions
\end{tabular} \\
\addlinespace
Adequate but requires significant improvements (3) &
\begin{tabular}[t]{@{}p{10cm}@{}}
$\bullet$ Workarounds can often be made to work, but require substantial manual effort \\
$\bullet$ Reasoning about mutable state is inherently difficult \\
$\bullet$ Extensions such as GSVerif can help address this gap
\end{tabular} \\
\addlinespace
Adequate with minor limitations (2) &
\begin{tabular}[t]{@{}p{10cm}@{}}
$\bullet$ Tables can handle state, though not always easily \\
$\bullet$ Debugging stateful models is difficult
\end{tabular} \\
\addlinespace
Fully adequate (2) & --- \\
\addlinespace
Not sure (2) & --- \\
\bottomrule
\end{tabular}
\end{table*}

\begin{table*}[t]
\centering
\caption{Participant ratings of ProVerif's support for reasoning about complex algebraic properties.}
\label{tab:q21}
\small
\renewcommand{\arraystretch}{1.3}
\begin{tabular}{@{}p{5cm} p{10cm}@{}}
\toprule
\textbf{Response} & \textbf{Representative concerns raised} \\
\midrule
Limited for algebra-heavy designs (4) &
\begin{tabular}[t]{@{}p{10cm}@{}}
$\bullet$ Almost no support for Diffie--Hellman (DH)-type equations or theories using AC operators, which are common in practice \\
$\bullet$ Does not support full DH reasoning the way Tamarin does
\end{tabular} \\
\addlinespace
Not sure (3) & --- \\
\addlinespace
Adequate with careful modeling (3) &
\begin{tabular}[t]{@{}p{10cm}@{}}
$\bullet$ Can handle Diffie--Hellman, but one must be careful about the equations written and account for properties like associativity and commutativity \\
$\bullet$ Expected to be of comparable quality to Tamarin
\end{tabular} \\
\addlinespace
Fully adequate for modern protocols (1) & --- \\
\addlinespace
Adequate but prone to false positives (1) &
\begin{tabular}[t]{@{}p{10cm}@{}}
$\bullet$ Termination is not obvious
\end{tabular} \\
\addlinespace
Insufficient and requires redesign (1) &
\begin{tabular}[t]{@{}p{10cm}@{}}
$\bullet$ No support for AC symbols in ProVerif
\end{tabular} \\
\bottomrule
\end{tabular}
\end{table*}

\begin{table*}[t]
\centering
\caption{Participant ratings of ProVerif's support for observational equivalence, with representative open-ended concerns for each category.}
\label{tab:q20}
\small
\renewcommand{\arraystretch}{1.3}
\begin{tabular}{@{}p{5cm} p{10cm}@{}}
\toprule
\textbf{Response} & \textbf{Representative concerns raised} \\
\midrule
Insufficient for modern protocols (4) &
\begin{tabular}[t]{@{}p{10cm}@{}}
$\bullet$ Diff-equivalence is too coarse or too strong, limiting precision for properties like anonymity or unlinkability \\
$\bullet$ Limited to reasoning about processes that differ only by certain terms \\
$\bullet$ Gaps in applicability to IoT and post-quantum computing settings
\end{tabular} \\
\addlinespace
Adequate but requires significant improvements (3) &
\begin{tabular}[t]{@{}p{10cm}@{}}
$\bullet$ Diff-equivalence is an over-approximation that is sometimes hard to overcome \\
$\bullet$ Termination issues \\
$\bullet$ Need for more specific explanations of failure causes
\end{tabular} \\
\addlinespace
Adequate with minor limitations (2) &
\begin{tabular}[t]{@{}p{10cm}@{}}
$\bullet$ Desire to examine more than two executions (e.g., for robust declassification) \\
$\bullet$ Sufficient, but usability requires deep prior knowledge of the tool
\end{tabular} \\
\addlinespace
Fully adequate (2) & --- \\
\addlinespace
Not sure (2) & --- \\
\bottomrule
\end{tabular}
\end{table*}


\begin{table*}[t]
\centering
\caption{Summary of requested usability, design, and functional improvements.}
\label{tab:requested-improvements}
\small

\begin{tabularx}{\textwidth}{
    >{\raggedright\arraybackslash}p{0.16\textwidth}
    >{\raggedright\arraybackslash}p{0.22\textwidth}
    >{\raggedright\arraybackslash}p{0.22\textwidth}
    >{\raggedright\arraybackslash}X
}
\toprule
\textbf{Improvement Area} &
\textbf{Tamarin Evidence} &
\textbf{ProVerif Evidence} &
\textbf{Main Implication} \\
\midrule

Error feedback and diagnostics &
Error feedback: 8(67\%) responses in Q19 &
Automated suggestions for modeling mistakes (3 responses); interactive debugging (8 responses) &
Errors should identify the relevant model element, explain the cause, and recommend possible corrective actions. \\
\midrule

Counterexample and result explanation &
Counterexample explanation: 8(67\%) responses; high-level AI synthesis requested in open-ended responses &
Graphical attack-trace exploration (6 responses) &
Verification outcomes should be explained at the protocol level and connected to assumptions and modeling decisions. \\
\midrule

Modeling and validation support &
Modeling support: 6(50\%) responses &
Model-validation mechanism (10 responses) &
Tools should assist model construction and help users assess whether the formal model matches the intended protocol. \\
\midrule

Interactive verification and debugging &
Better interactive verification and proof-oracle support requested in open-ended responses &
Interactive debugging mode (8 responses) &
Users need step-by-step access to proof states, bottlenecks, unresolved branches, and attacker behavior. \\
\midrule

Visualization &
Improved interactive and counterexample support implied by responses &
Graphical model and attack-trace interface (6 responses) &
Visualizations should map internal proof structures to protocol roles, messages, states, and properties. \\
\midrule

Documentation and examples &
Documentation: 5(42\%) responses &
Improved documentation(5 responses) &
Documentation should include realistic annotated models, modeling rationale, failure cases, and debugging workflows. \\
\midrule

Proof automation and guidance &
Better proof oracles and case-distinction support; AI synthesis requested by 2 respondents &
Helper-lemma generation requested by 2 respondents &
Tools could recommend helper lemmas, case distinctions, model simplifications, and proof strategies. \\
\midrule

Tooling and workflow support &
Tooling support: 5(42\%) responses &
IDE or CI/CD integration (3 responses) &
Editor and workflow integration is useful, but currently less important than interpretability and validation. \\
\midrule

Functional and expressive support &
Greater support needed for stateful protocols and observational equivalence &
Greater support needed for state, algebraic theories, observational equivalence, and state-evolving properties &
Some usability barriers require improvements to the tools' underlying verification and modeling capabilities. \\
\bottomrule
\end{tabularx}
\end{table*}
\begin{table*}[t]
\centering
\renewcommand{\arraystretch}{1}
\setlength{\tabcolsep}{6pt}

\begin{tabularx}{\textwidth}{
    |>{\centering\arraybackslash}m{0.09\textwidth}
    |>{\centering\arraybackslash}X
    |>{\centering\arraybackslash}m{0.14\textwidth}
    |>{\centering\arraybackslash}m{0.18\textwidth}
    |>{\centering\arraybackslash}m{0.19\textwidth}|
}
\hline
\textbf{ID} &
\textbf{Role} &
\textbf{Experience} &
\textbf{Expertise} &
\textbf{Tool(s)} \\
\hline



P1 & Academic Researcher or Faculty & 1--3 yrs & Intermediate &
ProVerif \\
\hline

P2 & Graduate & 1--3 yrs & Deep Expertise &
ProVerif, Squirrel \\
\hline

P3 & Academic Researcher or Faculty & 4--6 yrs & Deep Expertise &
Tamarin \\
\hline

P4 & Academic Researcher or Faculty & > 6 yrs & Deep Expertise &
Tamarin, ProVerif, CryptoVerif \\
\hline

P5 & Academic Researcher or Faculty & 4--6 yrs & Advanced &
Tamarin, EasyCrypt, ProofFrog \\
\hline

P6 & Industry Professional & > 6 yrs & Advanced &
Isabelle/PSPSP, OFMC \\
\hline

P7 & Academic Researcher or Faculty & > 6 yrs & Deep Expertise &
Tamarin, ProVerif, OFMC, AVISPA, AVANTSSAR, SPaCIoS \\
\hline

P8 & Academic Researcher or Faculty & > 6 yrs & Advanced &
Tamarin, ProVerif \\
\hline

P9 & Academic Researcher or Faculty & > 6 yrs & Advanced &
Tamarin, ProVerif, DeepSec \\
\hline

P10 & Graduate student & 1--3 yrs & Intermediate &
Tamarin, ProVerif, AVISPA \\
\hline

P11 & Academic Researcher or Faculty & 1--3 yrs & Intermediate &
ProVerif \\
\hline

P12 & Academic Researcher or Faculty & > 6 yrs & Deep Expertise &
Tamarin, ProVerif \\
\hline

P13 & Industry professional & > 6 yrs & Advanced &
Tamarin, ProVerif \\
\hline

P14 & Academic Researcher or Faculty & 4--6 yrs & Advanced &
Tamarin, ProVerif, DeepSec \\
\hline

P15 & Academic Researcher or Faculty & > 6 yrs & Advanced &
Tamarin, ProVerif \\
\hline

P16 & Academic Researcher or Faculty & > 6 yrs & Deep Expertise &
Tamarin, ProVerif \\
\hline

\end{tabularx}

\caption{Demographic details of survey participants.}
\label{tab:participant-demographics}
\end{table*}
\cleardoublepage

\end{document}